\documentclass[oldversion]{aa}  
\usepackage{graphicx,psfig,amsmath}

\begin{document}
   \title{Rayleigh scattering by H$_2$ in the extrasolar planet HD\,209458b}

   \author{
   A.~Lecavelier des Etangs\inst{1,2}
   \and
   A.~Vidal-Madjar\inst{1,2}
   \and 
   J.-M.~D\'esert\inst{1,2}
   \and 
   D.~Sing\inst{1,2}
}
   
   \authorrunning{}

   \offprints{A.L. (\email{lecaveli@iap.fr})}

   \institute{
   CNRS, UMR 7095, 
   Institut d'Astrophysique de Paris, 
   98$^{\rm bis}$ boulevard Arago, F-75014 Paris, France
   \and
   UPMC Univ. Paris 6, UMR 7095, 
   Institut d'Astrophysique de Paris, 
   98$^{\rm bis}$ boulevard Arago, F-75014 Paris, France
  }   
   \date{Received 3 March 2008/ Accepted 8 April 2008}
 
  \abstract
{
Transiting planets, such as HD\,209458b, offer a unique opportunity 
to scrutinize the planetary atmospheric content. 
Although molecular hydrogen is expected to be the main atmospheric constituent,
H$_2$ remains uncovered because of the lack of strong transition from 
near-ultraviolet to near-infrared.
Here we analyse the absorption spectrum of HD\,209458b obtained by Sing et al. (2008a) 
which provides a measurement of the absorption depth in the 3000-6200\,\AA\ wavelength range.  We show 
that the rise in absorption depth at short wavelengths can be interpreted as
Rayleigh scattering within the atmosphere of HD\,209458b. 
Since Rayleigh 
scattering traces the entire atmosphere, this detection enables a direct 
determination of the pressure-altitude relationship, which is required to determine 
the absolute fraction of other elements such as sodium. At the zero altitude defined by the 
absorption depth of 1.453\%, which corresponds to a planetary radius of 0.1205 times the 
stellar radius, we find a pressure of 33$\pm$5\,mbar. Using the variation 
of the Rayleigh scattering cross-section as a function of wavelength, 
we determine the temperature to be 2200$\pm$260\,K at 33\,mbar pressure. 
}


\titlerunning{Rayleigh scattering in HD\,209458b}

   \maketitle

\section{Introduction}
\label{Introduction}

Transiting planets such as HD\,209458b provide a unique opportunity 
to scrutinize their atmospheric content 
({\it e.g.}, Charbonneau et al.\ 2002, Vidal-Madjar et al. 2003, 2004, 2008; 
Ballester et al.\ 2007; Barman 2007). 
The theory of transmission spectroscopy was developed in detail
by the pioneering works of Saeger \& Sasselov (2000), Brown (2001), and 
Hubbard et al.\ (2001).
Primary transits can reveal minute quantities of gas through spectral 
absorption leading to larger apparent planet size, or larger absorption depth 
in the transit light curve, at wavelengths characteristic of the absorber.
Due to Rayleigh scattering, the main atmospheric constituent, H$_2$, 
is seen by means of the ramp-up of apparent planetary radius from 500 to 300 nm
(Hubbard et al.\ 2001).
Using transmission spectroscopy, several minor constituents of 
the atmosphere of HD\,209458b have already been uncovered, including sodium, atomic hydrogen, carbon and oxygen. 

The extrasolar planet HD\,209458b was observed during primary 
transit using the {\sl Hubble Space Telescope} STIS instrument, 
at low resolving power using the G430L and G750L spectrograph, and
at medium resolution using the G750M spectrograph. 
The data was analyzed as described in Balleter et al.\ (2006) 
and Sing et al.\ (2008a); special care was taken 
to correct for the limb-darkening effect. In this article, we  
use the absorption depth (AD) of the planet, which was measured  
as a function of wavelength 
from 3000\,\AA\ to $\sim$6200\,\AA .

In the whole paper, the zero altitude in the planet atmosphere 
is defined to be
the mean absorption depth of 1.453\%, measured by Knutson et al.\ (2007). 
This corresponds to a planetary radius of 0.1205 times the 
stellar radius. 

In Sect.\ref{First esimates}, we present first estimates of temperature
and pressure that can be obtained assuming that the Rayleigh 
scattering by molecular hydrogen is responsible for the observed
increase in absorption depth at short wavelengths.
In Sect.\ref{Fit of the spectrum}, we present a fit to the data 
including Na\,{\sc i} absorption lines. 
Discussion and conclusion can be found in Sect.~\ref{Discussion}
and~\ref{Conclusion}.

\section{First estimates}
\label{First esimates} 

\subsection{The HD\,209458b transit spectrum from 3000 to 6200\,\AA}

In a plot of the absorption depth as a function of wavelength, 
an increase in absorption depth is observed toward shortest  wavelengths below $\sim$5000\,\AA\ (Fig.~\ref{fig1}). 
Above 5000\,\AA , the absorption depth remains roughly constant up to 5500\,\AA , before rising again from 5500 to 6000\,\AA . This last increase in absorption 
depth, which has a maximum at about 5900\,\AA\ is interpreted to be due to a high abundance 
of sodium in the bottom atmosphere, below the Na$_2$S cloud level (Sing et al.\ 2008b). 

In the short wavelength side, the mean absorption depth is measured 
to be 1.444\%+/-0.002\% between 4500 and 5500\,\AA , while it is 
1.475\%$\pm$0.005\% between 3000 and 3900\,\AA . 
This significant variation of 0.030\% in absorption 
depth was already identified in the same dataset (Ballester et al.\ 2007). 
Ballester et al. (2007) proposed that the increase in absorption depth
\emph{below 4000\,\AA} could be due to absorption in the Balmer series 
by an optically thin layer of excited hydrogen atoms at high altitude 
in the planetary atmosphere. 
Here we propose another explanation: Rayleigh scattering.

\subsection{Rayleigh scattering}
 
Following the derivation of Lecavelier des Etangs et al.\ (2008), 
we define $\tau_{\rm eq}$ as the optical depth at the altitude $z_{\rm eq}$, 
such that a sharp occulting disk of radius $R_p+z_{\rm eq}$ produces the same 
absorption depth (AD) as the planet with a translucent atmosphere.
In other words, $\tau_{eq}$ is defined to be
$AD  =(R_p+z( \tau= \tau_{\rm eq}  ))^2/R_{\rm star}^2$,
where $R_p$ and $R_{\rm star}$ are the planet and star radii, respectively.
Numerical integration of absorption depth from model atmospheres shows that 
$\tau_{\rm eq}$ is quasi-constant; its value is $\sim$0.56 for a wide range of 
atmospheric scale height, $H$, provided $R_p/H$ 
is between $\sim$30 and $\sim$3000 (Lecavelier des Etangs et al.\ 2008). 
For a temperature $T$, the atmosphere scale height is given by $H=k T/\mu g$,
where $g$ is the gravity and $\mu$ is the mean molecular mass.
In the case of HD\,209458b, $R_p/H$ varies between 100 and 1000,  when the 
temperature varies from 300 to 3000\,K. For a given atmospheric structure 
and composition, the theoretical absorption depth at a wavelength $\lambda$ 
can be therefore calculated by finding $z(\tau=\tau_{\rm eq},\lambda)$, which solves 
the equation $\tau(z,\lambda)= \tau_{\rm eq}=0.56$
(for details, see Lecavelier et al.\ 2008).

The Rayleigh scattering cross-section 
follows a power-law function of wavelength to the power of four, of the form:  
$\sigma(\lambda)=\sigma_0(\lambda/\lambda_0)^{-4}$, 
where $\sigma_0$ is the Rayleigh scattering cross-section 
at a reference wavelength $\lambda_0$. 
The high value of exponent in the power law is the reason behind the Earth's blue sky. 
In this case, equations from Lecavelier des Etangs et al.\ (2008) 
that give the planet radius as a function of the wavelength  
can be solved analytically. For the absorption depth of Rayleigh scattering, 
this leads to the simple equation 
\begin{equation}
AD=AD_0 \left(1-\frac{8H}{R_p} \ln \frac{\lambda}{\lambda_0}\right) ,
\label{AD}
\end{equation}
where $AD_0$ is the absorption depth at a reference 
wavelength $\lambda_0$. 

Two important quantities appear in this equation. 
First, the absorption depth follows a linear relation as a 
function of the logarithm of wavelength; the slope is 
characteristic of the scale height H, which is directly 
proportional to the temperature. 
Second, the total density and therefore the pressure at 
the altitude defined by $AD_0$ corresponds to density and pressure
at which Rayleigh scattering is opaque at wavelength $\lambda_0$.
Thus, the determination of the couple ($AD_0$,$\lambda_0$) provides
an absolute reference for the altitude-pressure relationship.
This absolute reference is required to interpret measurements 
of relative absorption depth by minor species such as sodium, 
and to determine their abundances (Sing et al. 2008b). 

The temperature is directly related to the observed 
variation in absorption depth as a function of wavelength.
For HD\,209458b, assuming a temperature of 1500\,K 
that provides a typical scale height of 550\,km with the 
planetary parameters of Knutson et al.\ (2007), and using 
a mean molecular mass of $\mu$=2.3\,$m_p$, where $m_p$ is the mass 
of the proton, the variation in absorption depth is expected to be 
to be -0.034\% from 3000\,\AA\ to 5000\,\AA ;
this is similar to what is observed in the G430L data. 
In the other way, 
if interpreted in term of Rayleigh scattering, the observed 
increase in absorption depth at short wavelengths of 0.031\% 
between 3500 and 5000\,\AA\ (Sect.\ref{Introduction}) 
implies a temperature of about 2000\,K. 

Concerning the pressure, the fit of the measured absorption depth 
from 3200 to 5000\,\AA\ by the law given in Eq.~\ref{AD} 
provides $AD_0$ =1.453\% at 
$\lambda_0$=4300\,\AA , where 1.453\% corresponds to the mean planet 
radius measured by Knutson et al.\ (2007). This allows the derivation of the 
pressure at the planet radius (zero altitude) to be 
$P_0= \tau_{\rm eq}/ \sigma_0 \times \sqrt{kT \mu g/2 \pi R_p}$, 
where $\sigma_0$  is the Rayleigh scattering cross-section at 
4300\,\AA\ (Lecavelier des Etangs et al.\ 2008). 
Using the refractive index of molecular hydrogen $(r-1) =1.32\times 10^{-4}$, 
one derives $\sigma_0=2.3\times 10^{-27}$\,cm$^{2}$ at 
$\lambda_0$=4300\,\AA , and $P_0\sim$30\,millibars.

\section{Fit of the spectrum}
\label{Fit of the spectrum}

\subsection{Fit with Rayleigh scattering and sodium line}

\begin{figure}[tbh!]
\psfig{file=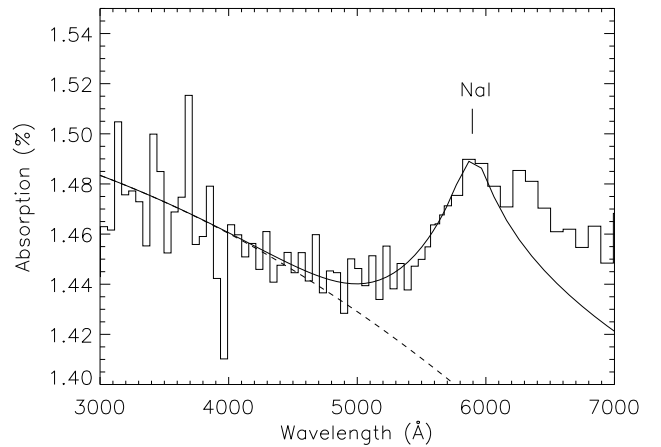,width=\columnwidth,angle=90} 
\caption[]{
Plot of absorption depth as a function of 
wavelength (histogram). 
The data were rebinned by 20~pixels 
corresponding to 55\,\AA\ per bin at wavelengths below 5500\,\AA . 
The measured absorption depths have typical 1-$\sigma$ error bars 
of 0.015\% to 0.020\% per bin below 4000\,\AA , and about 0.010\% 
above 4500\,\AA . 
The thick line shows the best fit to the data using the 
temperature-pressure-altitude profile given in Sing et al.\ (2008b). 
The dashed line shows the absorption with Rayleigh scattering only.
\label{fig1}}
\end{figure}

Beyond simple estimates using a model that includes only Rayleigh scattering, 
we can measure the atmospheric structure of HD\,209458b more accurately 
using a global fit to the data which introduces other absorbents, and 
includes medium-resolution data which further constrains 
the characteristics of sodium in the high atmosphere (Sing et al.\ 2008b). 
With other absorbents, the absorption depth cannot be calculated 
analytically and a numerical fit to the data is required. 

We fit the whole set of data, including data acquired using the 
medium resolution G750M spectrograph (Sing et al.\ 2008a, 2008b).
In the fit, the pressure and temperature profile as a function 
of altitude, as well as the sodium abundance, are free parameters. 
There is a total of 7~free parameters 
that are described in Sing et al.\ (2008b): five parameters describe 
the T-P profile and two parameters, the sodium abundance below 
and above the sodium condensation temperature. 
The sodium absorption line 
profile is calculated using collisional broadening (Iro et al.\ 2005).
We obtain a satisfactory fit, which has a $\chi^2$ of 40 for 52~degrees 
of freedom in the low-resolution data set (Fig.~\ref{fig1}).
The results for sodium clouds and the corresponding temperature-pressure 
diagram are given and discussed in Sing et al.\ (2008b). 

Defining zero altitude to be an absorption depth of 1.453\%, 
we find that the temperature and pressure at this altitude
are constrained mainly by Rayleigh scattering below 5000\,\AA . 
We find $P_0=33\pm 5$\,millibars and $T_0=2200\pm 260$\,K (1-$\sigma$).
At this pressure and temperature, 
the Rayleigh scattering makes the atmosphere optically 
thick at wavelength shorter than $\sim$5000\,\AA , and 
causes a significant increase in the absorption 
depth of $\sim$0.03\% from 5000\,\AA\ to 3000\,\AA\ (Fig.~\ref{fig1}).
The blue wing of the sodium line overcomes the absorption due to 
Rayleigh scattering at wavelengths longer than $\sim$5000\,\AA . 
An additional absorption in the red wing of the sodium line, 
beyond 6200\,\AA , could be explained by TiO and VO molecules 
(D\'esert et al., in preparation).
The sodium abundance required to fit the blue tail of the line profile 
is found to be $X({\rm NaI})= 4.3^{+2.1}_{-1.3}\times 10^{-6}$, 
which is two to four times the solar abundance (Asplund el al.\ 2005).

\begin{figure}[tbh!]
\psfig{file=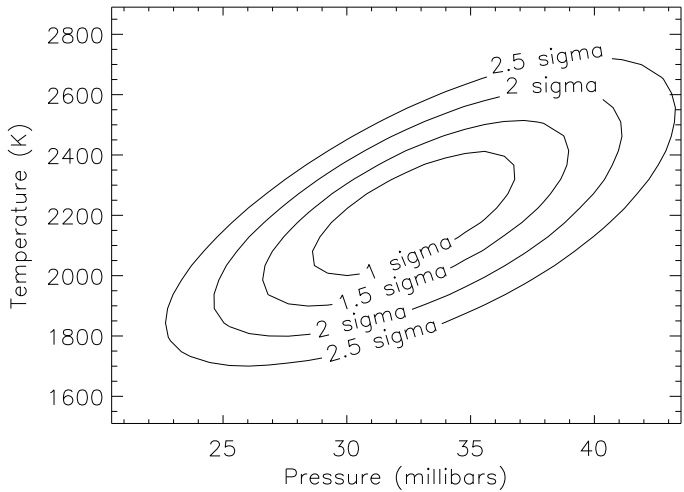,width=\columnwidth,angle=90} 
\caption[]{
Plot of error bars for the temperature and the pressure 
at the zero altitude level corresponding to the altitude 
at which the absorption depth is 1.453\%. 
\label{fig2}}
\end{figure}

Error bars for temperature and pressure 
at the zero altitude level, are plotted in Fig.\ref{fig2}.
These error bars are calculated using the $\chi^2$ difference 
between the data and a fit to the absorption depth measurements 
from 3200 to 6200\,\AA . 
For a given temperature and pressure at zero altitude, 
the other parameters of the fit are free to vary. 
Therefore, the resulting error bars include the uncertainties 
in the sodium abundance and other parameters of the T-P profile.

This fit to the data does not assume an isothermal
temperature profile as function of altitude in the atmosphere. Even more, the temperature gradient 
is found to be large and close to the limit allowed by the adiabatic gradient
(Sing et al.\ 2008b). An isothermal assumption is used only to 
calculate the column density in the limb grazing line of sight. 
Following Fortney (2005), we used an isothermal decrease of density 
with altitude to derive an approximation of the column 
density, $N_{iso}=n\sqrt{2\pi H R_p}$, where $n$ is the 
volume density at lowest altitude in the line of sight 
(see also Lecavelier des Etangs et al.\ 2008). 
With a temperature gradient $dT/dz$, a better approximation 
of the column density is given by 
$N\approx N_{\rm iso}\sqrt{Tdz/HdT}\,\Gamma(1/2+Tdz/HdT)/\Gamma(1+Tdz/HdT)$, 
that is $N\approx N_{\rm iso} (1-0.125 HdT/Tdz)$.
For the maximum temperature gradient that corresponds to the onset of convection
(adiabatic T-P profile) and is given by $HdT/Tdz=2/7$, 
the difference between $N_{\rm iso}$ and the actual column density 
is less than 4\%.
The results provided here are thus unaffected by the actual shape of 
the temperature profile.

\subsection{Alternative scenario}

\subsubsection{Balmer jump}

\begin{figure}[bth!]
\psfig{file=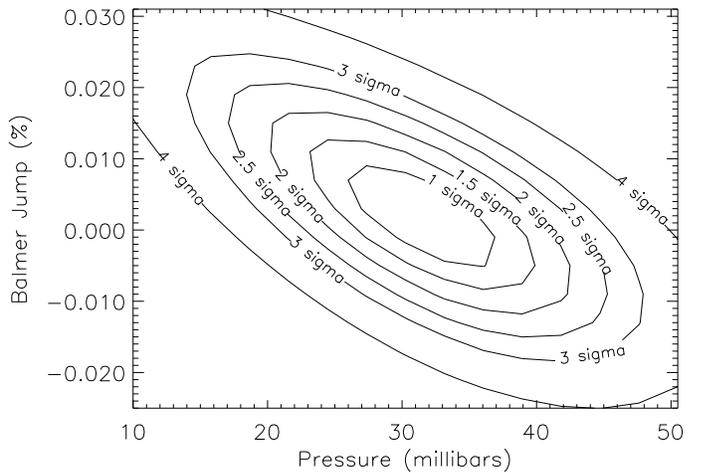,width=\columnwidth,angle=90} 
\caption[]{Plot of error bars for the pressure and 
the Balmer increment ($\Delta AD_{\rm Balmer}$). 
This shows that a sharp increase in absorption depths
at the limit of the Balmer series is not significantly detected. 
The increase in absorption depth of $\sim$0.030\% is 
more likely to be due to Rayleigh scattering by gas 
at $\sim$33\,mbar pressure than by a Balmer jump of $\sim$0.030\%.
\label{fig Balmer}}
\end{figure}

The rise of absorption depth at short wavelengths has already been 
interpreted in terms of an extended cloud 
of excited hydrogen that is optically thin at wavelengths shorter 
than the Balmer limit (Ballester et al.\ 2007).
To analyze this possibility, we add one free parameter to the model with a jump 
in absorption depth between 3600 and 3900\,\AA , $\Delta AD_{\rm Balmer}$. 
The additional absorption is taken to be zero above 3900\,\AA\ and 
linearly increases to reach $\Delta AD_{\rm Balmer}$ at 3600\,\AA\ 
and below. A fit to the data using this additional free parameter provides 
$\Delta AD_{\rm Balmer}$= 0.002\%$\pm$0.007\%.
The best fit corresponds to a $\chi^2$ of 40, while a Balmer
increment of $\sim$0.030\% as claimed by Ballester et al.\ (2007)
increases the $\chi^2$ to $\ga 50$ and decreases the pressure 
at zero altitude to $\sim$15\,millibars 
to avoid absorption by Rayleigh scattering
(Fig.~\ref{fig Balmer}).
We therefore conclude that Rayleigh scattering with a smooth
increase in absorption depth at short wavelengths better fit the 
observational data than the ``Balmer jump'' with its sharp increase 
in absorption depth below $\sim 4000$\,\AA .

Alternatively, the ``Balmer jump'' in absorption due to a cloud of excited 
hydrogen could also be consistent with the measurements, if we consider 
an additional absorber to explain the low variation in absorption depth
between 4000 and 5000\,\AA . 
Therefore, we add a second free parameter to the model, namely a minimum absorption depth, 
$AD_{\rm min}$, which is expected when 
an optically thick cloud layer is present at high altitude. 
For these two additional parameters, we find that  
$AD_{\rm min}$=1.4452$\pm$0.0016\% and $\Delta AD$= 0.031\%$\pm$0.007\%. 
In this model, fit to the data again constrains the pressure at zero altitude 
to be lower than $\sim$15\,millibars. Since the sodium line profile 
is pressure broadened, a lower pressure at reference altitude requires a far higher sodium
abundance to fit the blue tail of the absorption line profile.  
Using this lower limit for the pressure, we find that the sodium 
abundance must be $X({\rm NaI})=1.5\times 10^{-5}$ 
which is 10 times the solar abundance. 
In addition, the lower pressure at zero altitude
also constrains the pressure at which the core of 
the sodium line is detected to be below $\sim$$10^{-6}$\,bar; this is
a very low pressure at which all atomic sodium should 
have disappeared through ionization by UV flux, 
even at lower altitude (Fortney et al.\ 2003). 
A fit to the data using a cloud layer and a ``Balmer jump'' therefore 
requires a high sodium abundance at significant height
in the atmosphere where sodium is believed to be ionized.
The ``Balmer jump'' scenario thus appears to be unlikely.

We conclude that Rayleigh scattering is probably the 
simplest and most robust interpretation of the data 
because the absorption-depth curve from 3000 to 
6000\,\AA , including the rise between 4000 and 
5000\,\AA , can be interpreted by means of Rayleigh 
scattering and a sodium abundance close to solar. In 
contrast, the ``Balmer jump'' would require two additional 
components and a high sodium abundance.

\subsubsection{Atomic lines}

\begin{figure}[bth!]
\psfig{file=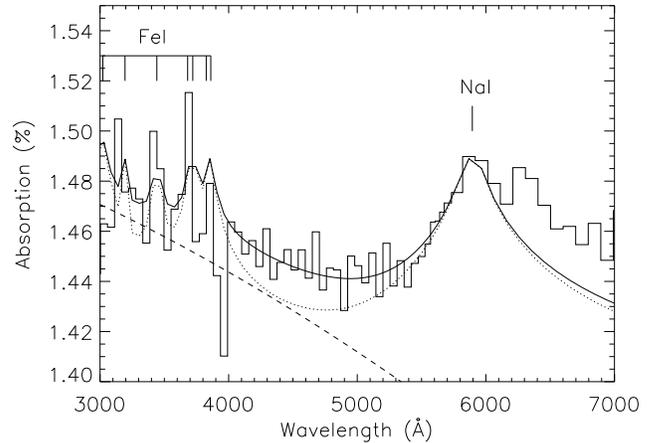,width=\columnwidth,angle=90} 
\caption[]{A similar plot to Fig.~\ref{fig1}, with fit including absorption 
by atomic iron lines. When the atomic iron lines are included, 
a satisfactory fit to the data can be obtained assuming a lower 
pressure of $\sim$15 millibars and larger abundance for sodium 
and iron of about 10 times solar abundances. 
The dotted line shows the absorption depth obtained neglecting 
the Rayleigh scattering. This demonstrates that even with large 
iron abundance, the Rayleigh scattering has a significant 
contribution to the absorption depth level between 4000 and 5000\,\AA , 
imposing a minimum pressure at the corresponding altitude. 
\label{fig3}}
\end{figure}

We considered another alternative to Rayleigh scattering 
by the addition of plausible atomic lines to interpret the increase 
in absorption depth at the shortest wavelengths below 5000\,\AA\ (Barman 2007). 
Taking into account the species abundances, the strongest atomic 
lines in the 3000-5000\,\AA\ wavelength range are those 
of Ca\,{\sc i}, Fe\,{\sc i}, Al\,{\sc i}, and Cr\,{\sc i}. 
Since the strong calcium line at 4230\,\AA\ is not observed, 
the abundance of atomic calcium must be extremely low. The same conclusion 
applies to aluminium at 3950\,\AA\ and to chromium at 4270 
and 3600\,\AA . These species cannot be responsible for the 
observed absorption at short wavelengths.

Nonetheless, atomic iron presents a series of 
strong lines from 2900 to 3860\,\AA .
When the atomic iron lines are included, 
a satisfactory fit to the data is achieved when we assume 
(i) a lower pressure of $\sim$15 millibars; 
(ii) a higher abundance of both sodium and iron, 
of about 10 times solar abundance, and 
(iii) very low calcium, aluminium, and chromium abundances (Fig.~\ref{fig3}).
The higher sodium abundance is required to fit the blue wing 
of the line profile, because the line is pressure-broadened, 
and the pressure is lower. 

Calculating the best 
fit solution with and without iron lines (Fig.~\ref{fig1} and~\ref{fig3}), 
we find similar differences between the data and fit with $\chi^2$=40 for 52 
degrees of freedom in both cases. This does not favor either one 
of the two models. However, the model without lines of atomic iron 
is preferred because it does not require the large ten times solar 
abundance for sodium and iron. As for calcium, 
chromium and magnesium, iron is supposed to be trapped in 
hydrogen compound such as FeH. Because iron has the highest 
condensation temperature, it is not expected to be present 
in the gas phase in the absence of other species.

\section{Discussion}
\label{Discussion}

It is noteworthy that the temperature is found to be relatively 
high compared to first estimates obtained from secondary transits 
measurements (Deming et al.\ 2005). 
However, a detailed model is needed to interpret the estimated 
brightness temperature obtained from secondary transits and
to determine the pressure and altitude at which this temperature
corresponds
(Burrows et al.\ 2005). The temperature obtained here, mainly constrained 
by the slope of the Rayleigh scattering, is consistent with temperature-pressure 
profiles obtained by some detailed models of the atmosphere of HD\,209458b 
which provides high temperatures up to 2000\,K at 100\,millibars, 
in the night side (Burrows et al.\ 2006). However, other models to interpret 
secondary transits measurements provide lower temperatures (Seager et al. 2005)
and lower temperature gradients in the millibars pressure domain
(Iro et al.\ 2005; Fortney et al.\ 2006). 
Nonetheless, a wide range of models are consistent with the data, and although efforts 
were made to consider the widest range possible (Seager et al.\ 2005), 
measurements remain limited and their interpretation 
is model-dependent, despite significant recent progress 
(Knutson et al. 2007; Burrows et al.\ 2007a).

As shown above, the temperature estimates, in the Rayleigh regime, 
relies mainly on the slope and, therefore, on the absorption depth measured 
in the 3000-4000\,\AA\ range, where noise is the highest. 
Nevertheless, a fit to only measurements for absorption depth above 
4000\,\AA\ provides also a high temperature of $T_0=2350\pm 300$\,K (1-$\sigma$)
at zero altitude. Therefore, although the temperature 
given in Sect.~\ref{Fit of the spectrum} 
must be considered in light of its large associated error bars, 
a high temperature is also required to explain the variation 
in absorption depth between 4000 and 5000\,\AA\ where error bars 
are the lowest. In addition, this is consistent with the temperature 
required to explain the high sodium abundance 
below the sodium condensation level (Sing et al. 2008b).

This fit, which uses only measurements above 4000\,\AA , 
strengthens the Rayleigh interpretation. 
The variation in absorption depth between 4000 and 5000\,\AA\ 
cannot be explained by the absorption in the 
Balmer series, while  
the extrapolation of the 4000-5000\,\AA\ 
measurements assuming Rayleigh scattering 
perfectly matches the absorption depth 
measured between 3000 and 4000\,\AA . 

Importantly, even if Rayleigh scattering is not detected and the observed 
increase in absorption depth is due to another phenomenon, 
such as the Balmer jump (Ballester et al.\ 2007) or 
atomic lines (Barman 2007), 
the present calculations provide a strong upper limit to the pressure 
at the absorption depth level of 1.453\%.  
The result of Sect.~\ref{Fit of the spectrum} shows that 
at this level, the pressure must be equal to or smaller than 
33~millibars, a pressure at which because of Rayleigh scattering
a grazing line of sight becomes optically thick at 4300\,\AA . 

The red wing of the magnesium line at 2850\,\AA\ could also 
explain the increase in absorption depth 
at the shortest wavelengths, however the abundances of both 
sodium and magnesium would have to exceed approximately 
$\sim$20~times solar.
We therefore conclude that the increase in absorption depth 
at the shortest wavelengths 
is best explained by Rayleigh scattering. This represents 
the first direct detection of the main constituent of the 
planetary atmosphere: molecular hydrogen.

\section{Conclusion}
\label{Conclusion}

Before, the determination of the planetary radius-pressure relationship 
was missing although this is needed to determine absolute abundances. 
Until now, many models have relied on the assumption that 
pressure at the mean planet transit radius is close to 1\,bar. 
Using Rayleigh scattering, we now determine that this pressure 
should be $\sim$30\,millibars. Even if the Rayleigh scattering does 
not dominate the absorption, the pressure must be lower than 
$\sim$30\,millibars. Consequently, the radius of HD\,209458b at 1~bar 
is determined to be about 1.29\,Jupiter radius, slightly lower 
than previously assumed. However, the difference is not sufficient 
to explain the puzzling large radius of HD\,209458b (Guillot et al.\ 2006; 
Burrows et al.\ 2007b; Chabrier et al.\ 2007). 

In the near future, the new generation of instruments with increased capabilities 
in the near UV should enable the physical properties of a constantly 
increasing list of transiting extrasolar planets to be determined. 
When Rayleigh scattering by H$_2$ in these planets is detected, it will provide the reference baseline 
for the determination of the absolute abundances of all the elements 
to be detected using absorption spectroscopy (Ehrenreich et al.\ 2006). 
Measurement of the atmospheric composition will become an increasingly 
important tool to help understand the origin and nature of 
extrasolar planets, as increasingly smaller extrasolar planets 
are detected (Gillon et al.\ 2007). 

\begin{acknowledgements}
We warmly thank Drs. G.~Ballester, F.~Bouchy, 
D.~Ehrenreich, R.~Ferlet and G.~H\'ebrard for enlightening discussions.
We also thank the referee, Dr.~J.~Fortney, for his constructive and fruitful 
remarks.
\end{acknowledgements}

\end{document}